\documentclass[10pt,conference]{IEEEtran}

\usepackage[dvips]{graphicx}
\usepackage[latin1]{inputenc}
\usepackage{amsmath,amsfonts} 
\usepackage{color}
\usepackage[english]{babel}

\begin{document}

\title{Analysis of Non-binary Hybrid LDPC Codes}

\author{
\authorblockN{Lucile Sassatelli and David Declercq}
\authorblockA{ETIS\\
ENSEA/UCP/CNRS UMR-8051 \\
95014 Cergy, FRANCE\\
\{sassatelli,declercq\}@ensea.fr}}

%

\maketitle
\footnotetext[1]{This work was supported by the French Armament Procurement Agency (DGA).}
\begin{abstract}
This paper is eligible for the student paper award.\\
In this paper, we analyse asymptotically a new class of LDPC codes called Non-binary Hybrid LDPC codes, which has 
been recently introduced in \cite{Sassatelli06}. We use density evolution techniques to derive a stability condition for hybrid LDPC codes, and prove their threshold behavior. We study this stability condition to conclude on asymptotic advantages of hybrid LDPC codes compared to their non-hybrid counterparts.

\end{abstract}

\section{Introduction}
Like Turbo Codes, LDPC codes are pseudo-random codes which are well-known to be channel capacity-approaching. LDPC codes have been rediscovered by MacKay under their binary form and soon after their non-binary counterpart have been studied by Davey \cite{Davey_IEEECommLetter1998}. Non-binary LDPC codes have recently received a 
great attention 
because they have better performance than binary LDPC codes for short block length and/or high order modulations \cite{Hu_ICC2004,PoulliatTC2006,Bennatan_IEEETransInfTheo2006}. However, good short length non-binary LDPC codes tend to be 'ultra-sparse', and have worse convergence threshold than binary LDPC codes.

Our main motivation in introducing and studying the new class of hybrid LDPC codes is to combine the 
advantages of both families of codes, binary and non-binary. 
Hybrid codes families aim at achieving this trade-off by mixing different order for the symbols in the 
same codeword. Our resulting codes are called Non-binary Hybrid LDPC Codes because of the mixture of 
different symbol sets in the codeword.

In \cite{Sassatelli06}, we have demonstrated the interest of the Hybrid LDPC codes by designing codes 
that compare favorably with existing codes for quite moderate code length (a few thousands bits). 
Hybrid LDPC codes appear to be especially interesting for low rate codes, $R\leq 0.25$. In this 
paper, we study the asymptotic behavior and properties of Hybrid LDPC codes under iterative belief 
propagation (BP) decoding.

The section two of this paper highlights the generality of our new codes structure, and explains why 
we have focused the asymptotic study on the particular subclass of linear codes. In third and fourth 
sections, we present the context of the study, and detail symmetry and linear-invariance properties 
which are useful for the stability condition.
This condition is then expressed 
and analyzed to show theoretic advantages of Hybrid LDPC codes.

\section{The Class of Hybrid Codes}
\begin{figure}[!h]
\centering
\input{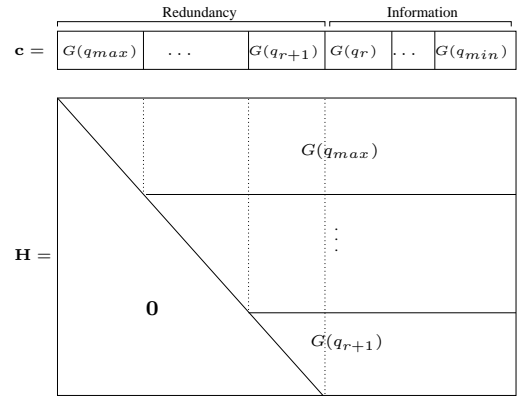}
\caption{Hybrid codeword and parity-check matrix.}
\label{H}
\end{figure}

We define a Non-binary Hybrid LDPC code as LDPC code whose variable nodes belong to finite sets of different orders. 
More precisely, this class of codes is not defined in a finite field, but in finite groups.
We will only consider groups whose cardinality $q_k$ is a power of $2$, that says groups of the type $G(q_k)=\left(\frac{\mathbb{Z}}{2\mathbb{Z}}\right)^{p_k}$ with
$p_k=\log_2(q_k)$. Thus to each element of $G(q_k)$ corresponds a binary map of $p_k$ bits. Let us call the minimum order of codeword symbols $q_{min}$, and the maximum order of codeword symbols $q_{max}$. The class of hybrid LDPC codes is defined on the product group $\left(\frac{\mathbb{Z}}{2\mathbb{Z}}\right)^{p_{min}}
\times\hdots\times\left(\frac{\mathbb{Z}}{2\mathbb{Z}}\right)^{p_{max}}$.
Let us notice that this type of LDPC codes built on product groups has already been proposed in the literature \cite{Sridhara_ITW2002}, but no optimization of the code structure has 
been proposed and its application was restricted to the mapping of the codeword symbols to different modulation orders. Parity check codes defined on $\left(G(q_{min})\times \ldots \times G(q_{max})\right)$ are particular since they are linear in $\frac{\mathbb{Z}}{2\mathbb{Z}}$, 
but could be non-linear in the product group. Although it is a loss of generality, we have decided to restrict ourselves to hybrid LDPC codes that 
are linear in their product group, in order to bypass the encoding problem. We will therefore only consider upper-triangular parity check matrices and a 
specific sort of the symbol orders in the codeword, which ensures the linearity of the hybrid codes. The structure of the codeword and the 
associated parity check matrix is depicted in Figure \ref{H}.
We hierarchically sort the different group orders in the rows of the parity-check matrix, and also in the codeword, such that 
$q_{min}<\ldots<q_{k}<\ldots<q_{max}$. To encode a redundancy symbol, we consider each symbol that participates in the parity check 
as an element of the highest group, which is only possible if the groups are sorted as in Figure \ref{H}. This clearly shows that encoding 
is feasible in linear time by backward computation of the check symbols.\\
In order to explain the decoding algorithm for hybrid LDPC codes, it is useful to interpret a parity check of a hybrid code as a special case of a parity check 
built on the highest order group of the symbols of the row, denoted $G(q_l)$ and have a look at the binary image of the equivalent code \cite{PoulliatTC2006}. For codes defined 
over Galois fields, the nonzero values of $\mathbf{H}$ correspond to the companion matrices of the finite field elements and are typically rotation matrices (because of the 
cyclic property of the Galois fields). 

In the case of hybrid LDPC codes, a nonzero value is a function that connects a row in $G(q_l)$ and a column in $G(q_k)$, i.e., that maps the $q_k$ symbols of $G(q_k)$ into a subset of $q_k$ symbols that belongs to $G(q_l)$. Such application is not necessarily linear, but in the case it is, its equivalent binary representation is a matrix of dimension $(p_l \times p_k)$. Note that, with the above mentioned constraints, we have necessarily $p_k\leq p_l$. It is possible to generalize the 
Belief propagation decoder to hybrid codes, and it has been shown that even for those very specific structures, it is possible to derive a fast version of the decoder using FFTs \cite{Goupil_IEEETransCommun2006}. In this work, we consider only maps that are linear applications, and hence that have a binary representation, in order to be able to apply all known results on linear codes. We call the message passing step through $h_{ij}$ 
(cf. figure \ref{fig2}) \textit{extension} when it is from $G(q_k)$ to $G(q_l)$ and \textit{truncation} when it is from $G(q_l)$ to $G(q_k)$.
\begin{figure}[!h]
\input{hybridcheck.pstex_t} 
\caption{Parity-check of an hybrid LDPC code.}
\label{fig2}
\vspace*{-0.5cm}
\end{figure}

\section{Properties of Linear Hybrid LDPC Codes}
\subsection{The Extension and Truncation Operations}
We first clarify the nature of the non-zero elements of the parity-check matrix of a hybrid LDPC code.
We consider an element $A$ of the set of linear \textit{extension}s from $G(q_k)$ to $G(q_l)$. 
Im($A$) denotes the image of $A$.
$A$ belongs to the set of linear applications from $G(2)^{p_k}$ to $G(2)^{p_l}$ which are full-rank (that is injective since dim(Im($A$))=rank($A$)=$p_k$).
\begin{eqnarray}
A:\qquad G(2)^{p_k}&\rightarrow& G(2)^{p_l}\nonumber\\
\mathbf{i}&\rightarrow&\mathbf{j}\nonumber
\end{eqnarray}
$\mathbf{i}$ denotes the binary map of $i$ in $G(q_k)$ in $G(2)^{p_k}$, with $p_k=\log_2(q_k)$. That is, each index $i$ is taken to mean the $i$th element of $G(q_k)$, given some enumeration of the field. $x_i$ is the $i$th element of vector $\mathbf{x}$.
The \textit{extension} $\mathbf{y}$, of the probability vector $\mathbf{x}$ by $A$, is denoted by $\mathbf{x}^{\times A}$ and defined by: for all $i=0,\hdots,q_l-1$
\begin{eqnarray}
&\mbox{if }i\notin \mbox{Im}(A),&\mbox{ }y_i=0\nonumber\\
&\mbox{if }i\in \mbox{Im}(A),&\mbox{ }y_i=x_j\mbox{ with $j$ such that }\mathbf{i}=A\mathbf{j}\nonumber
\end{eqnarray}

$A$ is called \textit{extension}, and the inverse function $A^{-1}$ \textit{truncation} from Im($A$) to $G(q_k)$. The \textit{truncation} is defined by

\begin{eqnarray}
A^{-1}:\qquad \mbox{Im}(A)&\rightarrow& G(2)^{p_k}\nonumber\\
\mathbf{j}&\rightarrow&\mathbf{i}\mbox{ with }j\mbox{ such that }\mathbf{j}=A\mathbf{i}\nonumber
\end{eqnarray}
The \textit{truncation} $\mathbf{x}$ of the probability vector $\mathbf{y}$ by $A^{-1}$ is denoted by $\mathbf{y}^{\times A^{-1}}$ and defined by
\begin{eqnarray}
i=0,\hdots,q_k-1\mbox{, } x_i=y_j\mbox{ with $j$ such that }\mathbf{j}=A\mathbf{i}\nonumber
\end{eqnarray}

Given a probability-vector $\mathbf{x}$ of size $q$, the components of the logarithmic density ratio (LDR) vector $\mathbf{w}$ associated with $\mathbf{x}$ are defined as $w_i=\log \left(\frac{x_0}{x_i}\right),\mbox{ }i=0,\hdots,q-1$.\\
At channel output, LDR messages are actually logarithmic likelihood ratio (LLR) vectors.

\subsection{Parameterization of Hybrid LDPC family}

An edge of the Tanner graph of an Hybrid LDPC code has four parameters $(i,q_k,j,q_l)$. A hybrid LDPC code is then represented by $\pi(i,j,k,l)$ which is the proportion of edges connecting variable nodes of degree $i$ in $G(q_k)$, to check nodes of degree $j$ in $G(q_l)$. Thus, hybrid LDPC codes have a very rich parameterization since the parameter space has four dimensions.

\subsection{Symmetry definition for density evolution approach}

Let $\mathbf{W}$ be a LLR vector computed at the output of a discrete memoryless channel, 
and $v$ the component of the codeword sent, corresponding to the required value for the data node the edge with message $\mathbf{W}$ is connected to. 
$\mathbf{W}^a$ denotes the cyclic-permutation of $\mathbf{W}$. $c$ denotes the value of the symbol linked to the edge with the message $\mathbf{W}$, and $y$ the available information on all other edges of the graph. $W_i^a$ is the $i$th component of $\mathbf{W}^a$ and is defined by $W_i^a=\log \left(\frac{P(a\cdot c=0|y)}{P(a\cdot c=i|y)}\right)$, where $\cdot$ denotes the multiplication in $G(q)$.
Like in \cite{Bennatan_IEEETransInfTheo2006}, $\mathbf{W}^a$, for all $a\in G(q)$ is defined by
\[W_i^a=W_{a+i}-W_a,\quad \forall i=0\dots q-1\]
A channel is cyclic if output LLR vector $\mathbf{W}$ fulfills
\[P(\mathbf{W}^a|v=0)=P(\mathbf{W}|v=a)\]
\underline{Definition 1} On a cyclic channel, a LDR message is symmetric, if the following expression holds
\[\forall a\in G(q),\quad P(\mathbf{W}=\mathbf{w}|v=a)=e^{-w_a}P(\mathbf{W}=\mathbf{w}|v=0)\]
Most practical channels are cyclic, and thus, in this work, we assume transmission on arbitrary memoryless cyclic-symmetric channels. The generalization of the results in this paper to non-symmetric channels can be done thanks to the coset approach as in \cite{Bennatan_IEEETransInfTheo2006}. The symmetry property is the essential condition for any asymptotic study since it ensures that the error probability is independent of the codeword sent.\\
\underline{Lemma 1}: If $\mathbf{W}$ is a symmetric LDR-vector random variable, then its \textit{extension} $\mathbf{W}^{\times A}$, by any linear \textit{extension} $A$ with full rank, is also symmetric.\\
The same lemma holds for \textit{truncation}. 
The data pass and the check pass of belief propagation have already been shown to preserve symmetry. Thus, lemma 1 
ensures that the hybrid decoder preserves the symmetry property if the input messages are symmetric.\\
\underline{Lemma 2}: The error-probability of a code in a hybrid family, used on a cyclic-symmetric channel, is independent on the codeword sent.\\
For lack of space, we do not give the proof of this lemma, which is a direct generalization of \cite{Urbanke2001}.

\subsection{Linear Application-Invariance}
Now we introduce a property that is specific to the hybrid codes families.
Bennatan et al. in \cite{Bennatan_IEEETransInfTheo2006} used permutation-invariance to derive a stability condition for non-binary LDPC codes, and to approximate the densities of graph messages using one-dimensional functionals, for extrinsic information transfert (EXIT) charts analysis.
The difference between non-binary and Hybrid LDPC codes holds in the non-zeros elements of the parity-check matrix. Indeed, they do not correspond anymore to cyclic permutations, but to linear \textit{extensions} or \textit{truncations}, that we denote by linear applications.
The goal is to prove that linear application-invariance (shortened by LA-invariance) of messages is induced by choosing uniformly the linear \textit{extensions} which are the non-zero elements of the hybrid parity-check matrix. In particular, LA-invariance allows to characterize message densities with only one scalar parameter \cite{Sassatelli06}.\\
We work with probability vector random variables, but all the definitions and proofs given in the remaining also apply to LDR-vector random variables. We denote by $E$ the set of linear \textit{extension}s from $G(q_1)$ to $G(q_2)$, and by $T$ the set of "inverse functions" of $E$, what we call the set of linear \textit{truncation}s from $G(q_2)$ to $G(q_1)$ (see previous section on linear \textit{extensions}).\\
\underline{Definition 2}: $\mathbf{Y}$ is LA-invariant if and only if for all $(A^{-1},B^{-1})\in T\times T$, the probability-vector random variables $\mathbf{Y}^{\times A^{-1}}$ and $\mathbf{Y}^{\times B^{-1}}$ are identically distributed.\\
\underline{Lemma 3}: If a probability-vector random variable $\mathbf{Y}$ of size $q_2$ is LA-invariant, then for all $(i,j)\in G(q_2)\times G(q_2)$, the random variables $Y_i$ and $Y_j$ are identically distributed.\\
\underline{Definition 3}: Let $\mathbf{X}$ be a $q_1$-sized probability-vector random variable, we define the \textit{random-extension} of size $q_2$ of $\mathbf{X}$, denoted $\tilde{\mathbf{X}}$, as the probability-vector random variable $\mathbf{X}^{\times A}$, where $A$ is uniformly chosen in $E$ and independent on $\mathbf{X}$.\\
\underline{Lemma 4}: A probability-vector random variable $\mathbf{Y}$ is LA-invariant if and only if there exists a probability-vector random variable $\mathbf{X}$ such that $\mathbf{Y}=\tilde{\mathbf{X}}$.\\
For lack of space reason, we will detail the proof of this lemma, which is easy, in a future publication. Thanks to lemma 4, the check node incoming messages are LA-invariant in the code family made of all the possible cycle-free interleavers and uniformly chosen linear \textit{extension}s (and hence corresponding \textit{truncation}s). Moreover, random-\textit{truncations}, at check node output, ensures LA-invariance of variable node incoming messages. Thus, as shown in \cite{Sassatelli06}
under Gaussian approximation, the densities of vector messages are characterized by only one parameter.

\section{The Stability Condition for Hybrid LDPC Codes}

The stability condition, introduced in \cite{Urbanke2001}, is a necessary and sufficient condition for the error probability to approach arbitrarily close to zero, assuming it has already dropped below some value at some iteration. In this paragraph, we generalize the stability condition to hybrid LDPC codes.\\
Given a hybrid family defined by $\pi(i,j,k,l)$, we define the following family parameter:
\begin{eqnarray}
\Omega&=&\sum_{j,k,l} \pi(i=2,k,j,l) \frac{q_k-1}{q_l-1}(j-1)\nonumber
\end{eqnarray}
Also for a given memoryless symmetric output channel with transition probabilities $p(y|x)$ and a mapping $\delta(\cdot)$, we define the following channel parameter:
\[\Delta=\sum_{k,l} \pi(k,l) \frac{1}{q_l-1} \sum_{i=1}^{q_k-1} \int \sqrt{p(y|\delta(i))p(y|\delta(0))}dy\]
E.g., for BI-AWGN channel, we have
\[\Delta=\sum_{l,k} \pi(l,k)\frac{1}{q_l-1}\sum_{i=1}^{q_k-1}exp(-\frac{1}{2\sigma^2}n_i)\]
where $n_i$ is the number of ones in the binary map of $i\in G(q_k)$.\\
\underline{Theorem}: Let assume $(\pi,\delta)$ given for a hybrid LDPC set. Let $P_0$ denotes the probability distribution function of initial messages $\mathbf{R}^{(k)^{(0)}}$ for all $k$. Let $P_e^t=P_e(\mathbf{R}_t)$ denotes the average error probability at iteration $t$ under density evolution.
\begin{itemize}
\item If $\Omega\geq\frac{1}{\Delta}$, then there exists a positive constant $\xi=\xi(\pi,P_0)$ such that $P_e^t>\xi$ for all iterations $t$.
\item If $\Omega<\frac{1}{\Delta}$, then there exists a positive constant $\xi=\xi(\pi,P_0)$ such that if $P_e^t<\xi$ at some iteration $t$, then $P_e^t$ approaches zero as $t$ approaches infinity.
\end{itemize}

For lack of space reason, we give there only a sketch of the proof.\\
\underline{Proof}
\begin{itemize}
\item We first give the general lines of the proof of the necessary condition. Let $\mathbf{R}_{t+n}^{(k)}$ denotes the variable node outcoming messages in $G(q_k)$ at iteration $t+n$, where $n=0,1,\hdots$. Since we consider only cyclic-symmetric channels, we can apply lemma 4 from \cite{Bennatan_IEEETransInfTheo2006}. It ensures that there exists an erasurized channel such that the cyclic-symmetric channel is a degraded version of it, and hence provides a lower bound on the error probability. Let $\hat{\mathbf{R}}_{t+n}^{(k)}$, $n=0,1,\hdots$, denote the respective messages of the erasurized channel, and $\hat{\epsilon}_0$ the erasure probability. In the remainder of the proof, we switch to log-density representation of messages. Let $\hat{\mathbf{R^\prime}}_{t+n}^{(k)}$ denote the LDR-vector representation of $\hat{\mathbf{R}}_{t+n}^{(k)}$, $n=0,1,\hdots$. $Q_n^{(k)}(\mathbf{w})$ denotes the distribution of $\hat{\mathbf{R^\prime}}_{t+n}^{(k)}$. $P_0^{(k)}$ denotes the distribution of the initial message $\mathbf{R^\prime}_0^{(k)}$ of the cyclic-symmetric channel. The overline notation $\overline{\mathbf{X}}$ applied to vector $\mathbf{X}$ represents the vector resulting from random \textit{extension} followed by random \textit{truncation} of $\mathbf{X}$. Provided that random \textit{extension} and \textit{truncation} are such that $\overline{\mathbf{X}}$ and $\mathbf{X}$ are of same size, we can show that the error probabilities are equal. Thus, if $\overline{\mathbf{Q}_n^{(k)}}$ is the distribution of $\overline{\hat{\mathbf{R^\prime}}_{t+n}^{(k)}}$, we have
\[P_e(\mathbf{Q}_n)=\sum_k \pi(k) P_e(\mathbf{Q}_n^{(k)})=\sum_k \pi(k) P_e(\overline{\mathbf{Q}_n^{(k)}})\]
Therefore $P_e(\mathbf{Q}_n)$ is lower bounded by a constant strictly greater than zero if and only if there exists $k$ such that $P_e(\overline{\mathbf{Q}_n^{(k)}})$ is lower bounded by a constant strictly greater than zero.
Defining
\[\Omega_k=\sum_{j\geq 2,l} \pi(i=2,j,l|k) \frac{q_k-1}{q_l-1}(j-1)\]
and $\qquad\overline{P_0}=\sum_k \pi(k) \overline{P_0^{(k)}}$,
we show that
\begin{equation}\label{eqgeq}
P_e(\overline{\mathbf{Q}_n^{(k)}})\geq \frac{1}{2(q_{max}-1)^2} \hat{\epsilon}_0^{(k)^2} \Omega_k \Omega^{n-1} \overline{P_0}^{n-1}
\end{equation}
We prove that $\overline{P_0}$ is symmetric in the binary sense, and as in \cite{Bennatan_IEEETransInfTheo2006}, we obtain
\[\lim_{n\rightarrow \infty} \frac{1}{n} \log \overline{P_0}(W_1\leq 0)^{\otimes n}=\log\left(\mathbb{E}\left(-\frac{1}{2}R_1^{\prime}\right)\right)\]
where $R_1^{\prime}$ is the shortened notation for the first component of the mixture of decoder input LLR-vector random variables $\mathbf{R^\prime}_0^{(k)}$. We have
\[\mathbb{E}\left(-\frac{1}{2}R_1^{\prime}\right)=\mathbb{E}_{A,B}\left(\mathbb{E}\left(\sqrt{\frac{R_1^{\times A\times B^{-1}}}{R_0^{\times A\times B^{-1}}}}|A,B\right)\right)\]
and finally obtain
\[\mathbb{E}\left(-\frac{1}{2}R_1^{\prime}\right)=\sum_{k,l} \pi(k,l) \frac{1}{q_l-1} \sum_{i=1}^{q_k-1} E\left(\sqrt{\frac{R_i}{R_0}}\right)\]
and
\[E\left(\sqrt{\frac{R_i}{R_0}}\right)=\int\sqrt{p(y|\delta(i)p(y|\delta(0)))}dy\]
Hence, we find $\mathbb{E}\left(-\frac{1}{2}R_1^{\prime}\right)=\Delta$. This last equation combined with equation (\ref{eqgeq}) leads to the conclusion that $P_e(\overline{\mathbf{Q}_n^{(k)}})$ is lower bouded by a strictly positive constant, as $n$ tends to infinity, as soon as $\Omega\Delta \geq 1$. This condition is the same for all $k$. Thus, the necessary condition for stability is $\Omega<\frac{1}{\Delta}$.

\item We give now the main steps for the proof of the sufficiency of the condition.
$\mathbf{X}^{(k)}$ denotes a probability-vector random variable of size $k$. We define $D_n$ and $D_a$:
\begin{eqnarray}
D_n(\mathbf{X}^{(k)})&=&\mathbb{E}\left(\sqrt{\frac{\overline{X_i^{(k)}}}{\overline{X_0^{(k)}}}}\right)=\mathbb{E}\left(\sqrt{\frac{\overline{X_1^{(k)}}}{\overline{X_0^{(k)}}}}\right)\nonumber\\
&=&\sum_l \pi(l|k)\frac{1}{q_l-1}\sum_{i=1}^{q_k-1}\mathbb{E}\left(\sqrt{\frac{X_i^{(k)}}{X_0^{(k)}}}\right)\nonumber\\
D_a(\mathbf{X}^{(l)})&=&\frac{1}{q_l-1}\sum_{j=1}^{q_l-1} \mathbb{E}\left(\sqrt{\frac{X_j^{(l)}}{X_0^{(l)}}}\right)\nonumber
\end{eqnarray}
To shorten the notations we can omit the index of iteration $t$. The data pass is translated by
\[R_i^{(k)}=R_i^{(0)^{(k)}}\prod_{n=1}^{i-1}L_i^{(k)}\]
We obtain
{\footnotesize
\begin{eqnarray}
D_n(\mathbf{R}_t)&=&\mathbb{E}_{i,k}\left(\sqrt{\frac{\overline{R_i^{(k)}}}{\overline{R_0^{(k)}}}}\right)=\Delta \sum_k \pi(k) \left[\mathbb{E}\left(\sqrt{\frac{L_i^{(k)}}{L_0^{(k)}}}\right)\right]^{i-1}\nonumber
\end{eqnarray}}
First, we are going to prove the recursive inequality (\ref{rec})
\begin{figure*}
{\small
\begin{equation}\label{rec}
D_n(\mathbf{R}_{t+1})\leq \Delta \sum_{i,k} \pi(i,k) \left[\sum_l \pi(l|i,k) \frac{q_k-1}{q_l-1}\left(1-\sum_{j\geq 2} \pi(j|i,k,l)(1-\beta_t D_n(\mathbf{R}_t))^{j-1}\right)\right]^{i-1}+O(D_n(\mathbf{R}_t)^2)
\end{equation}
}
\end{figure*}
We show the three following equations.
\[\mathbb{E}\left(\sqrt{\frac{L_i^{(k)}}{L_0^{(k)}}}\right)=D_n(\mathbf{L}^{(k)})\]
\[D_n(\mathbf{L}^{(k)})=\sum_l \pi(l|k)\frac{q_k-1}{q_l-1} D_a(\mathbf{L}^{(l)})\]
\begin{eqnarray}
1-D_a(\mathbf{L}^{(l)})\geq &\sum_j \pi(j|l) (1-D_a(\mathbf{R}^{(l)}))^{j-1}&\nonumber\\
&+O(D_a(\mathbf{R}^{(l)})^2)&\nonumber
\end{eqnarray}
Connecting $D_a(\mathbf{R}^{(l)})$ to $D_n(\mathbf{R}_t)$ ends up with the proof of equation \ref{rec}:
\[D_a(\mathbf{R}^{(l)})=\frac{1}{q_l-1}\sum_k \pi(k|l) \sum_{i=1}^{q_k-1}\mathbb{E}\left(\sqrt{\frac{R_i^{(k)}}{R_0^{(k)}}}\right)\]
\[D_n(\mathbf{R}_t)=\sum_l \pi(l)\frac{1}{q_l-1}\sum_k \pi(k|l) \sum_{i=1}^{q_k-1}\mathbb{E}\left(\sqrt{\frac{R_i^{(k)}}{R_0^{(k)}}}\right)\]
we obtain 
\[D_n(\mathbf{R}_t)=\sum_l \pi(l)D_a(\mathbf{R}^{(l)})\]
We express $D_a(\mathbf{R}^{(l)})$ in terms of $D_n(\mathbf{R}_t)$:
\begin{eqnarray}
1-D_a(\mathbf{R}^{(l)})&\leq& 1-\min_l D_a(\mathbf{R}^{(l)})\leq 1-\beta_t D_n(\mathbf{R}_t)\nonumber
\end{eqnarray}
as soon as $\beta_t$ is a function of the iteration such that $\beta_t\leq \frac{\min_l D_a(\mathbf{R}^{(l)})}{D_n(\mathbf{R}_t)}$. Thus, we obtain equation (\ref{rec}).\\
We then can prove that if $\Omega< \frac{1}{\Delta}$ then there exists $\alpha$ such that if $D_n(\mathbf{R}_{t_0})<\alpha$ at some iteration $t_0$, then $\lim_{t\rightarrow \infty} D_n(\mathbf{R}_t)=0$. Moreover, if $\mathbf{X}^{(k)}$ is a symmetric probability-vector random variables of size $q_k$, then
\begin{equation}\label{lem}
\frac{1}{q_k^2}D_n(\mathbf{X}^{(k)})^2\leq P_e(\mathbf{X}^{(k)})\leq (q_k-1) D_n(\mathbf{X}^{(k)})
\end{equation}
Let us remind that $D_n(\mathbf{R}_t)=\sum_k \pi(k) D_n(\mathbf{R}_t^{(k)})$, and that the sequence $D_n(\mathbf{R}_t)_{t=t_0}^\infty$ converges to zero. Thus for all $k$, the sequences $D_n(\mathbf{R}_t^{(k)})_{t=t_0}^\infty$ also converge to zero.
And hence $P_e(\mathbf{R}_{t}^{(k)})$ converges to zero. This is true for all $k$, and since we have $P_e(\mathbf{R}_{t})=\sum_k\pi(k) P_e(\mathbf{R}_{t}^{(k)})$, $P_e(\mathbf{R}_{t})$ also converges to zero. This proves the sufficiency of the stability condition.
\end{itemize}\begin{flushright}$\square$\end{flushright}
Thus, we have proved that, provided that a fixed point of density evolution exists for hybrid codes, this point can be stable under certain condition. Our hybrid codes have hence threshold behavior.\\

\section{Analysis of the Stability Condition}
\begin{figure}
\begin{minipage}{0.5\textwidth}
      \begin{minipage}{.45\textwidth}
\includegraphics[width=4.3cm]{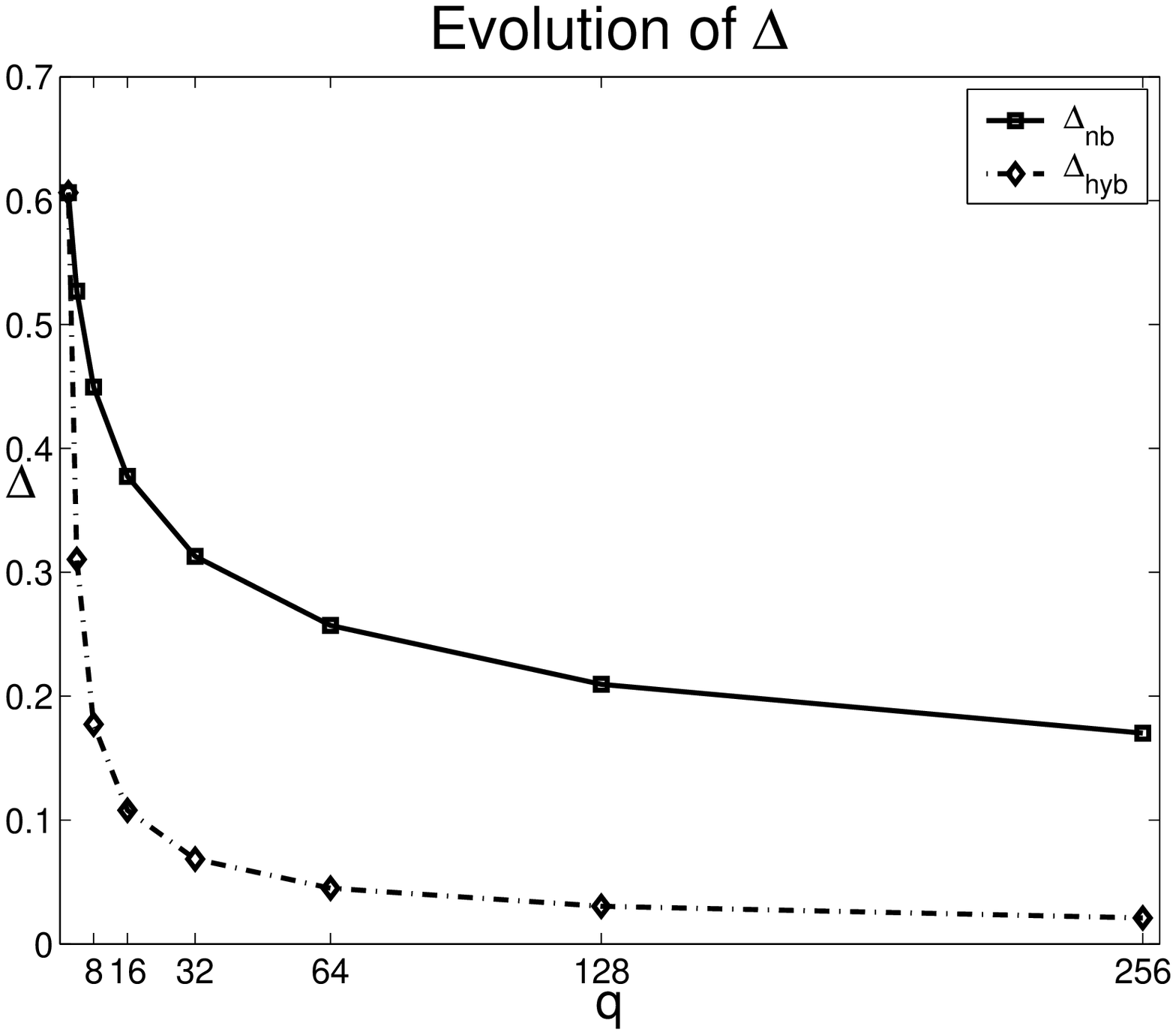}
		\end{minipage}
      \hfill                              
      \begin{minipage}{.5\textwidth}
\includegraphics[width=4.3cm]{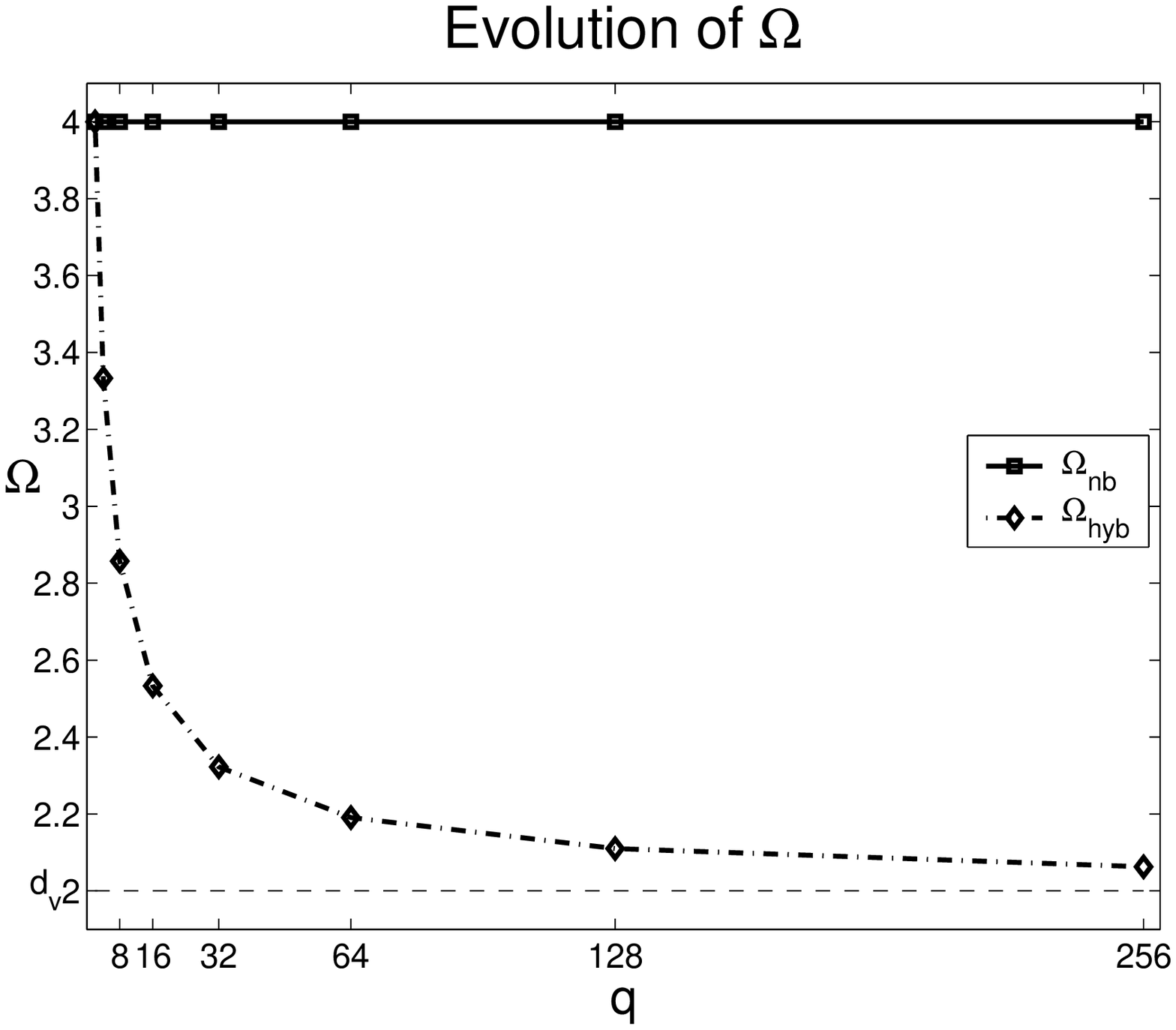}
      \end{minipage}
\end{minipage}
\caption{Channel and code parameters $\Delta$ and $\Omega$ for hybrid and non-hybrid codes in terms of maximum symbol order $q$. These figures show that a hybrid code can be stable when a non-binary code is not.}
\label{fig}
\end{figure}

Now we are able to compare the stability conditions for hybrid LDPC codes whose highest order group is $G(q)$ and for non-binary LDPC codes defined on the highest field $GF(q)$.
To illustrate advantages of hybrid codes over non-binary codes concerning the stability, we consider on figure \ref{fig} a code rate of one half, achieved by non-binary codes on $GF(q)$, with $q=2\dots 256$, and hybrid codes of type $G(2)-G(q)$, hence with graph rate varying with $q$. The information part of hybrid LDPC codes is in $G(2)$, and the redundancy in $G(q)$. We assume regular Tanner graphs for those codes, with connection degree of variable nodes $d_v=2$. The connection degree of check nodes will be hence varying whith the graph rate for hybrid codes.
We consider BI-AWGN channel whose variance $\sigma^2$ is set to $1$. We denote by $\Omega_{nb}$ and $\Omega_{hyb}$ the parameters of $GF(q)$ LDPC codes and hybrid LDPC codes, respectively. The same for $\Delta_{nb}$ and $\Delta_{hyb}$.\\
\underline{Remark $1$}: We note, on figure \ref{fig}, that $\Omega_{hyb}\leq \Omega_{nb}$ and $\Delta_{hyb}\leq \Delta_{nb}$ . Hence, with those assumptions, a fixed point of density evolution is stable at lower SNR for hybrid codes than for $GF(q)$ codes.\\
\underline{Remark $2$}: For a usual non-binary $GF(q)$ LDPC code, the hybrid stability condition reduces to non-hybrid stability condition, given by:
\begin{eqnarray}
\Omega_{nb}&=&\rho^\prime(1)\lambda^\prime(0)\nonumber\\
\Delta_{nb}&=&\frac{1}{q-1}\sum_{i=1}^{q-1}exp(-\frac{1}{2\sigma^2}n_i)\nonumber
\end{eqnarray}
with $n_i$, the number of ones in the binary map of $i\in G(q)$. Under this form, we can prove that $\Delta_{nb}$ tends to zero as $q$ goes to infinity. On BI-AWGN channel, this means that any fixed point of density evolution is stable as $q$ tends to infinity for non-binary LDPC codes, and for hybrid codes too (because of Remark $1$).\\
Those results indicate that optimization procedures will be more efficient since there exist more stable hybrid codes than non-hybrid LDPC codes for a given set of channel and code parameters. The optimization and code design is reported in a future application.

{\footnotesize
\bibliographystyle{unsrt}

}

\end{document}